\documentclass[10pt,letterpaper]{article}
\usepackage{JouSty}
\usepackage{cite}
\usepackage{hyperref}
\usepackage{booktabs}
\usepackage{graphicx}
\usepackage{amssymb,amsmath}
\usepackage{pgfplots}
\usepackage{epstopdf}
\DeclareMathAlphabet{\mathpzc}{OT1}{pzc}{m}{it}
\usepackage{multirow}
\usepackage{array}

\DeclareMathOperator{\sech}{sech}

\begin{document}
\title{Supercontinuum optimization for dual-soliton based light sources using genetic algorithms in a Grid platform}

\author{F. R. Arteaga-Sierra,${}^{1,2,*}$ C. Mili\'an,${}^{2,3,6}$ I. Torres-G\'omez,${}^{1}$ \\M. Torres-Cisneros,${}^{4}$ 
G. Molt\'o,${}^{3}$ and A. Ferrando${}^{2,5}$}

\address{
${}^{1}$Centro de Investigaciones en \'Optica, A.C., Le\'on Gto. 37150, M\'exico\\
${}^{2}$Universidad Polit\'ecnica de Valencia, Grupo de Modelizaci\'on Multidisciplinar Intertech, 46022, Valencia, Spain\\
${}^{3}$ Instituto de Instrumentaci\'on para Imagen Molecular (I3M). Centro mixto CSIC-Universitat Polit\`ecnica de Val\`encia-CIEMAT, camino de Vera s/n, 46022 Valencia, Espa\~na\\
${}^{4}$NanoBioPhotonics Group, DICIS, Universidad de Guanajuato, Mexico\\ 
${}^{5}$ Departament d'\`{O}ptica, Universitat de Val\`{e}ncia, Burjassot (Val\`{e}ncia). 46100, Spain\\ 
${}^{6}$Current address: Centre de Physique Th\'eorique, \'Ecole Polytechnique, CNRS, F-91128 Palaiseau, France
}
\email{*arteaga@cio.mx} 
\email{${}^6$carles.milian@cpht.polytechnique.fr} 

\begin{abstract}
We present a numerical strategy to design fiber based dual pulse light sources exhibiting two predefined spectral peaks in the anomalous group velocity dispersion regime. The frequency conversion is based on the soliton fission and soliton self-frequency shift occurring during supercontinuum generation. The optimization process is carried out by a genetic algorithm that provides the optimum input pulse parameters: wavelength, temporal width and peak power. This algorithm is implemented in a Grid platform in order to take advantage of distributed computing. These results are useful for optical coherence tomography applications where bell-shaped pulses located in the second near-infrared window are needed.
\end{abstract}


%%%%%%%%%%%%%%%%%%%%%%%%%%  body  %%%%%%%%%%%%%%%%%%%%%%%%%%
%%%% INTRODUCTION
%%%%%%%%%%%%%
\section{Introduction}

The Soliton self-frequency shift (SSFS) \cite{GO86,MITSCHKE86} plays a central role in many effects taking place during supercontinuum (SC) generation in optical fibers (see Refs. \cite{DU06,SK10} for a review on the topic). To mention only a few examples, light trapping \cite{GorbNP}, multi-peak soliton states \cite{HAUSE10,HM10,TR10}, emission of Airy waves \cite{Gorb08}, intense dark-soliton SC \cite{Milian09} or broad and intense blue shifted polychromatic dispersive waves \cite{MF12,Arteaga2014}, would not be possible (or strong enough) without the SSFS. One of the most notorious feature of the Raman effect in SC generation with femtosecond pulses corresponds to the Raman soliton carrying the lowest frequency. Its large frequency shift from the laser pulse has motivated infra-red (IR) Raman soliton sources \cite{DEK11,Roth12,ALK12} and their optimization \cite{JUD09,PRICK10,EGG10,GM10,Fe10}.

In the previous work of Ref. \cite{Arteaga2014}, we have shown that fs-pulses traveling in a dispersion engineered single mode fiber (SMF) can generate several pre-defined spectral peaks in the normal group velocity dispersion (GVD) region. These peaks correspond to the narrow band Cherenkov radiation emitted by the bright solitons \cite{AK95} undergoing Raman red-shift and recoil. Such spectra were important for applications in optical coherence tomography (OCT) with wavelengths in the near infra-red (NIR) window \cite{Wa02,Wa03,HUMB06}: $\lambda\lesssim1\ \mu$m.

In the present work, we are interested in OCT applications enabled by fiber based illumination \cite{Wang03,SP07} in the second near IR window (NIR II) \cite{Smith09,Hun10,Cao13} where $\lambda\in[1,1.4]\ \mu$m, typically in the anomalous GVD regime of highly nonlinear photonic crystal fibers (PCFs). Because of the typical dispersion landscape, sources in the NIR II window may be based on the bright optical solitons arising during SC generation through the intricate soliton fission effect \cite{KOyHA86,DribenPRA2013}. With this picture in mind, we used an efficient and general computational optimization method based on genetic algorithms (GA) \cite{SD10} that is capable of finding output spectra [see Fig. \ref{fig:Dispersion}] exhibiting two peaks centered at pre-defined wavelengths. This type of multi-peak spectral illumination is very often required for OCT applications in the NIR II \cite{Cao13} and the simultaneous presence of the two operating spectral components is a prerequisite for real time imaging \cite{Feldchtein98,Gelikonov04}. The two peaks we obtain are the first and second Raman solitons presenting a clean bell-shaped spectral profile, essential for OCT \cite{Fujimoto00}, with widths providing a decent longitudinal resolution $l_c\approx10\ \mu$m \cite{Fujimoto03}. To demonstrate the usefulness of this method, we consider in this work the two spectral channels separated by $100$ nm (see, e.g., Ref. \cite{Cao13}). The optimization method finds the optimal input pulse parameters, namely central wavelength, $\lambda_0$, temporal width, $T_0$, and peak power, $P_0$, yielding the desired spectra. The obtained peak powers are of up to $90$ mW for each spectral band, satisfying the needs for OCT imaging applications \cite{Fujimoto00}. We find by this method the possibility to tune the wavelength of the target spectral channels, what represents an important feature of this strategy since the greatest potential of the given PCF is exploited, specially in situations where limited choice of PCF designs is available.

It is worth mentioning that the use of GAs in optics has indeed proofed useful previously in solving satisfactorily the inverse optimization problem of the one we present here, i.e., the design of PCFs to control SC dynamics, in a wide range of situations \cite{KB04,TM10,TM09,RM07,YG11}. The use of GAs generally requires a large amount of simulations. For this reason we used a distributed computing (Grid) platform to reduce the time required to find the optimal solutions. The advantage of this infrastructure is that it enables the use of  the same code in a platform of scalable resources which are adapted according to the needs of the particular problem.
%%%%%%%%%%%%%
%%%% SC & GA
%%%%%%%%%%%%%
\section{Supercontinuum modeling and the Genetic Algorythm\label{sec:SCandGRID}}

We simulate the nonlinear propagation of the complex electric field envelope, $A$, along the fiber axis, $z$, by integrating numerically (with split-step fourier method) the generalized nonlinear Schr\"odinger equation (GNLSE) \cite{AG07},
% EQ 1
\begin{equation}
 -i\partial_z A(z,t) = \sum_{q \geq 2}\frac{\beta_q(\omega_0)}{q!}[i\partial_t]^q A(z,t)
 +\gamma A(z,t)\int^{+\infty}_{-\infty}dt'R(t')|A(z,t-t')|^2,
\label{eq:GNLSE}
\end{equation}
where the $\beta_q$'s account for the linear fiber dispersion and
% EQ 2
\begin{equation}
\gamma=\frac{\epsilon\epsilon_0^2\omega_0c}{3}\frac{\int\int dxdy n_2(x,y)[2|\vec{E}|^4+|\vec{E}^2|^2]}{[\int\int dxdy Re\lbrace \vec{E} \times \vec{H}^*\rbrace \hat{u}_z]^2},
\end{equation}
is the nonlinearity parameter \cite{AsMo09} which has been computed with a FEM solver (comsol) by integrating the electromagnetic components of the modal field at $\lambda=800$ nm along the transverse fiber cross section. $\epsilon=2.09$ and $n_2=2.6\times10^{-20}$ m$^2$/W are the relative permitivitty and nonlinear index of silica glass, respectively. The nonlinear response of the glass is $R(t)=[1-f_R]\delta(t)+f_Rh_R(t)$ \cite{Stolen:89}, where $\delta(t)$ is the Dirac delta function and the Raman (delayed) contribution is weighted with $f_R=0.18$ and described by the damped oscillator
% EQ 3
\begin{equation}
h_R(t)=\frac{\tau_1^2+\tau_2^2}{\tau_1\tau_2^2}\Theta(t)\mathrm{e}^{-\frac{t}{\tau_2}}\sin \left(  \frac{t}{\tau_1}\right) 
\label{eq:hR},
\end{equation}
where $\tau_1=12.2$ fs, $\tau_2=32$ fs, and $\Theta(t)$ is the Heaviside step function. The input pulses used in our simulations are of the form $\sqrt{P_0}\sech(t/T_0)$, where (for a fixed $\lambda_0$) the input peak power controls the soliton order $N\equiv T_{0} \sqrt{\gamma P_{0}/|\beta_{2}|}$.

\begin{figure}[htpb]
\centering
\includegraphics[width=6.5cm]{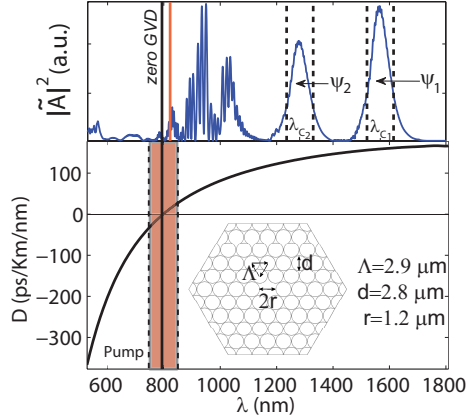}
\caption{\label{fig:Dispersion} (Bottom) Dispersion and cross section of the PCF used in our modeling. Shaded area marks the range of the input pump, $\lambda_0$. (Top) Typical spectral output and target channels centered at $\lambda_{c_{1,2}}$(delimited by dashed lines) in which $\psi_{1,2}$ are evaluated [See Eq. (4)]. The vertical black line marks the zero GVD and the red one the $\lambda_0$ used in this example.}
\end{figure}

\begin{figure}[htpb]
\centering
\includegraphics[scale=0.7]{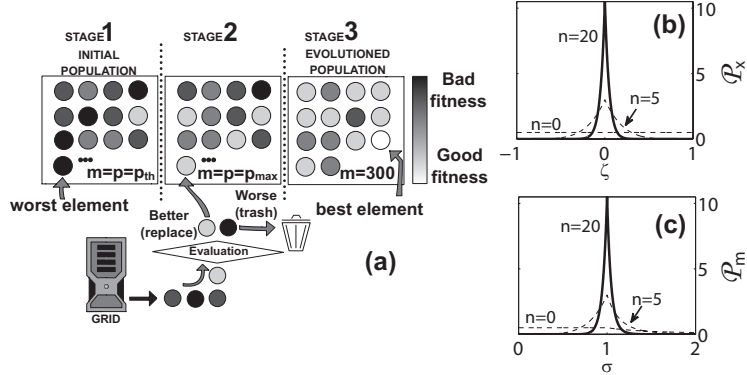}
\caption{\label{fig:geneticProb}(a) Sketch of the three GA stages: $p_{th}=50$, $p_{max}=100$. Probability distributions for (b) simulated binary crossover (SBX) and (c) polynomial mutation.}
\end{figure}

We consider a length $L=25$ cm of the NL-2.4-800 PCF, exhibiting the lowest zero GVD at $798$ nm [see details in Fig. \ref{fig:Dispersion}], convenient for broad band SC generation with Ti:Sapphire lasers. The frequency conversion performance is investigated within the attainable ranges of input pulse parameters: $\lambda\in[750,850]$ nm, $P\in[5,15]$ kW, and $T\in[30,150]$ fs.
% FIG 1
For each simulation along the PCF, the GA generates an \textit{individual} with the \textit{genome} $\mid g\rangle\equiv[g_1,g_2,g_3]^{T}=[T_0,\lambda_0,P_0]^{T}$ and evaluates how suitable that individual is from the simulation output through the \textit{fitness} function (to be minimized) defined as
% EQ 4
\begin{equation}
\label{eq:phi}
\phi\equiv\psi_1^{-1} \cdot \psi_2^{-1},\ \ \ \ \ \ \ \  \psi_{j}(\omega_{c_j};\Delta\omega)\equiv\int_{\omega_{c_j}-\Delta\omega}^{\omega_{c_j}+\Delta\omega}d\omega'|\tilde{A}(L,\omega')|^2,\ \ \ \ \ \ \ \  j=1,2,
\end{equation}
where $2\Delta\omega$ is the chosen spectral channel widths and $\omega_{c_j}=2\pi c/\lambda_{c_j}$ the central frequency. Note that the definition of $\phi$ as a product tends to favor output spectra in the form $\psi_1\approx\psi_2$ amongst all solutions with $\psi_1+\psi_2=const.$ [see Fig. (3) for optimum results]. The optimization process is depicted in Fig. \ref{fig:geneticProb}(a). 
% FIG 2
After an initial set of $p_{th}$ randomly (uniformly distributed) generated individuals (stage $1$), the genetic operators (GOs)  crossover, $\hat{\mathpzc{X}}$, and mutation, $\hat{\mathpzc{M}}$, are responsible for generating the new \textit{offsprings}, which are added to the population until the maximum size $p=p_{max}>p_{th}$ is reached (stage 2). During stage 3 the population size is kept constant,  $p=p_{max}$, and a \textit{replace the worst} strategy is used, i.e., the offspring is added to the population if $\phi_{offspring}<\phi_{max}$ or disregarded otherwise. This steady state GA keeps the Grid constantly computing new individuals in parallel and fully exploits the processing power of the Grid. We briefly describe the GOs below.

Cross-over, $\hat{\mathpzc{X}}$, is the first GO applied to the current population and generates two \textit{childs} $\mid g_c^{1,2}\rangle$ by combining two randomly chosen \textit{parents}, $\mid g_p^{1,2}\rangle$, i.e., $\hat{\mathpzc{X}}[\mid g_p^1\rangle^{T},\mid g_p^2\rangle^{T},\mid\emptyset\rangle^{T},\mid\emptyset\rangle^{T}]^{T}=[\mid g_p^1\rangle^{T},\mid g_p^2\rangle^{T},\mid g_c^1\rangle^{T},\mid g_c^2\rangle^{T}]^{T}$ (being $\mid\emptyset\rangle$ the empty set). We have used simulated binary crossover (SBX) \cite{Agrawal94} and the $[12\times12]$ operator
% EQ 5
\begin{equation}
 \hat{\mathpzc{X}}\equiv\left[\begin{array}{cccc}\hat{\mathpzc{I}} & \hat{0} & \hat{0} & \hat{0}\\ \hat{0} & \hat{\mathpzc{I}} & \hat{0} & \hat{0} \\ \hat{\alpha}_+ & \hat{\alpha}_- & \hat{0} & \hat{0} \\ \hat{\alpha}_- & \hat{\alpha}_+ & \hat{0} & \hat{0}\end{array}\right];\ \left(\hat{\alpha}_\pm\right)_{jk}\equiv x_k\frac{1\pm\bar{\sigma}_k}{2}\zeta_{jk},;\ x_k\equiv\Theta(u_k-0.05),\label{eq:Xoperator}
\end{equation}
where $\hat{0}\equiv0\times\hat{\mathpzc{I}}$ (being $\hat{\mathpzc{I}}$ the $2\times2$ identity matrix). The crossover \textit{activators}, $x_k$, set a probability for cross over of $95\%$ per gene. The stochastic variables in this case, $\bar{\sigma}_k$ [see Fig. \ref{fig:geneticProb}(b)], are chosen from a uniform random number $u_k\in[0,1]$ according to,
% EQ 6
\begin{equation}
\bar{\sigma}_k=\sigma\ /\int_{0}^\sigma{\mathpzc{P}}_x(\sigma)=u_k;\ {\mathpzc{P}}_x(\sigma)=\left\{\begin{array}{c}
0.5[n+1]\sigma^n,\ \sigma\leq1 \\ 0.5[n+1]\sigma^{-[n+2]},\ \sigma>1\end{array}\right.\label{eq:Px}.
\end{equation}
The polynomial mutation, $\hat{\mathpzc{M}}$, \cite{Deb2001} suitable for real coded problems (continuous valued variables), is applied after $\hat{\mathpzc{X}}$ and generates new genes as $\hat{\mathpzc{M}}:\mid g\rangle\rightarrow\mid g'\rangle$,
% EQ 7
\begin{equation}
 g_k'=g_k+m_k\Delta_k\bar{\zeta}_k;\ \ \ \ m_k\equiv\Theta \left( u_k-\frac{2}{3}\right)   \label{eq:Moperator},
\end{equation}
where $2\Delta_k$ is the interval size of each variable ($\Delta_\tau=60$ fs, $\Delta_\lambda=50$ nm, $\Delta P=5$ kW). Hence, in average only one gene is mutated per individual when mutation is applied: $\Theta(u_k-2/3)\Rightarrow P(m_k=1)=1/3$. $\bar{\zeta}_k\in[-1,1]$ satisfies the normalized probability distribution [see Fig. \ref{fig:geneticProb}(c)]
% EQ 8
\begin{equation}
{\mathpzc{P}}_m(\zeta)=0.5\left\{n+1[1-|\zeta|]^n\right\},
\end{equation}
with $n=20$, peaked around $\zeta=0$ and clearly different from the random generation. For each gene the stochastic variable $\bar{\zeta}$ is chosen from a random $u\in[0,1]$:
% EQ 9
\begin{equation}
\bar{\zeta}_k=\zeta /\int_{-1}^\zeta{\mathpzc{P}}_m(\zeta)=u_k;\ \ \ \ u_k\in[0,1].
\end{equation}
Statistically, $\hat{\mathpzc{M}}$ provides diversity to the population and $\hat{\mathpzc{X}}$ explores the parameter space in the vicinity of the parents. In this particular optimization problem, each individual evaluation typically required $\sim90$ s what amounted for about $7.5$ h of CPU time to perform a single run of the GA with $300$ individuals [see Fig. \ref{fig:CloudConv1}]. We used a cluster of $12$ cores within a Grid infrastructure, which reduced the computation time to $\sim40$ minutes. The Grid protocols supporting the GA execution make infrastructure resizable according to the needs of the problem: number of executions, dimensionality of the search space, etc. (see Ref. \cite{GM10} for details on the Grid).
%%%%%%%%%%%%%
%%%%     DUAL PEAK
%%%%%%%%%%%%%
\section{Dual-pulse solitonic source optimization\label{sec:Segunda-Parte}}
\begin{figure}%[htpb]
\begin{center}
\includegraphics[width=12cm]{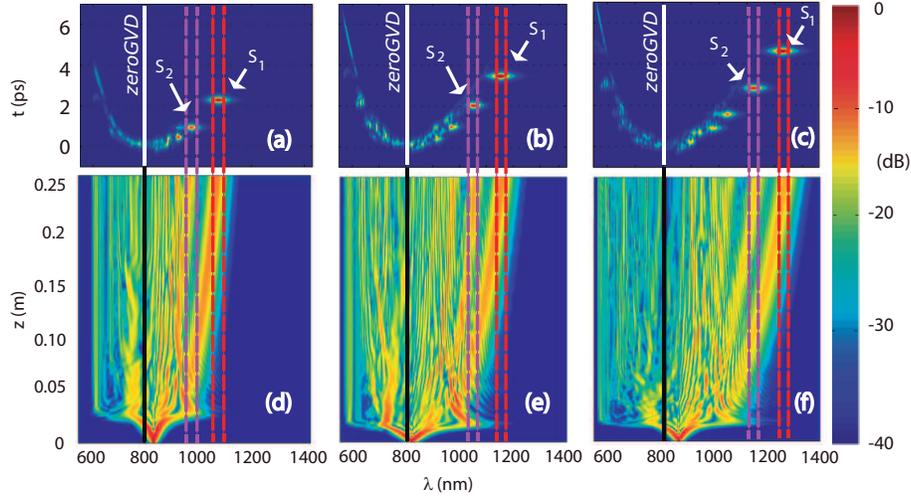}
\end{center}
\caption{\label{fig:NewEvoA}(a-c) Spectrograms of the output spectra, $z=25$ cm, corresponding to the optimization results given by the GA algorithm after $m=300$ evaluations. $S_{1,2}$ label the two solitonic pulses. (d-f) Spectral evolutions along the fiber associated to (a-c), respectively, retrieved from the optimal input pulse parameters corresponding to three different pairs of channels, $\lambda_{c_{1,2}}$. Input pulse parameters are given in Table I. Vertical solid lines mark the zero GVD wavelength.}
\end{figure}

As mentioned above, the desired output spectral channels in this work intend to cover OCT applications in the NIR II window, where transparency of the biological tissues increases and scattering decreases \cite{Bashkatov2005}. Because spectral bell-shaped pulses avoid spurious structures in OCT images \cite{Tripathi2002}, the bright optical solitons are very good candidates for OCT applications. Figure 3 shows the spectral evolutions (bottom) and output spectrograms (top) corresponding to the best individuals obtained by the GA strategy and fitness function, Eq. 4, described in the previous section. All output spectra shown in Fig. \ref{fig:NewEvoA} present the two reddest solitonic pulses, ejected from the soliton fission, accurately centered in the predefined channels ($\lambda_{c_{1,2}}$) delimited by the dashed lines (see Table I for parameter values associated to results in Fig. \ref{fig:NewEvoA}). In Figs. \ref{fig:NewEvoA}(a) and \ref{fig:NewEvoA}(d), the target spectral channels where chosen from Ref. \cite{Cao13} in order to illustrate the solution for a dual-pulse source required in a realistic application. The other two cases, Figs. 3(b,e) and Figs. \ref{fig:NewEvoA}(c,f), demonstrate the tunability of such source, keeping $\lambda_{c_{1}}-\lambda_{c_{2}}$ fixed to $100$ nm without replacing the PCF but merely adjusting the input pulse parameters. We checked by benchmarks that several runs of the GA with fixed  $\lambda_{c_{1,2}}$ provided systematically very similar optimal results and therefore only one is shown here for each different case.

Regarding OCT applications, another important aspect of the source presented here is that the fs-SC dynamics typically exhibits a very high coherence and negligible shot-to-shot fluctuations \cite{DU06}, known to be detrimental for OCT \cite{Fujimoto00}. Moreover, the two output solitonic pulses ($S_{1,2}$ in Fig. \ref{fig:NewEvoA}) constituting the proposed OCT light source, provide a decent resolution $l_c\equiv2\ln2\lambda_s/[\pi\Delta\lambda_{s,FWHM}]$ \cite{Izatt} of $\sim10\ \mu$m for the two solitons, $S_{1,2}$.
% Fig 3; 13.3 cm MAX
% Tabla 1 %
\begin{table}[ht]
\small
\begin{center}
\caption{\label{Table1} Parameters associated to the best individuals found by the GA, shown in Fig. \ref{fig:NewEvoA}.}
\begin{tabular}{c c c c l l c c}
  \hline
    \multicolumn{4}{c}{Optimal pulse parameters}     &        \multicolumn{1}{c}{Spectral bands} & \multicolumn{1}{c}{Resolution} & \multicolumn{1}{c}{Fitness} &       \multicolumn{1}{c}{Shown in}\\
  \cmidrule{1-4} \cmidrule(l){5-5} \cmidrule(l){6-6} \cmidrule(l){7-7} \cmidrule(l){8-8}
    $T_0$ (fs) & $\lambda_0$ (nm) & $P_0$ (kW) & N & $\lambda{c_1}$, $\lambda{c_2}$ (nm) & $l_{c_1}$, $l_{c_2}$ ($\mu$m) & $\phi$ ($10^{-4}/W$)  \\
  \hline
   90.01 & 834.98  & 5.012 & 20.14 & 1075, 975 & 9.8, \ \ 9.5 & 1.138 & Figs. 3(a,d)\\
70.80 & 817.27 & 12.3501 & 36.24 & 1150, 1050 & 10.6, 9.2 & 1.107 & Figs. 3(b,e)\\
101.13 & 849.24 & 9.6617 & 26.43 & 1225, 1125 & 10.1, 10.1 & 1.018 & Figs. 3(c,f)\\
  \hline
\end{tabular}
\end{center}
\end{table} 

\begin{figure}%[htpb]
\centering
\includegraphics[scale=.7]{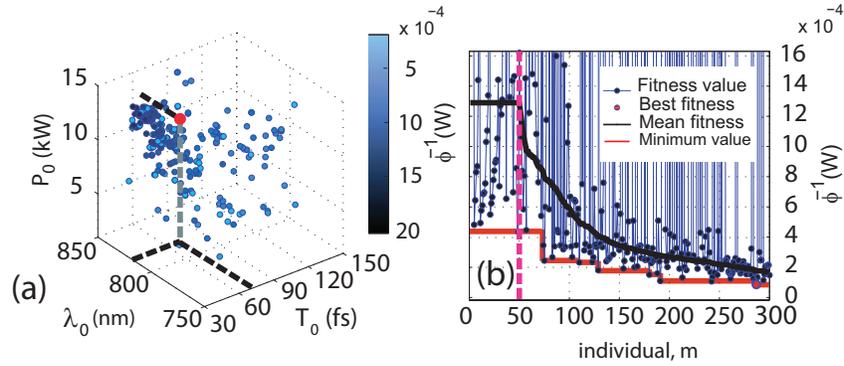}
\caption{\label{fig:CloudConv1}(a) Parameter space cloud of the $300$ individuals (and fitness) generated by the GA in the optimization yielding to the solution in Figs. 3 (b),(e). The best individual is marked in red and dashed mark its input parameters. (b) Fitness evolution versus generated individuals in chronological order. Dashed vertical line marks the threshold population $p_{th}=50$ corresponding to the end of stage 1 (random generation). Best (at $m\approx260$), Instantaneous minimum, and average fitness are also plotted (see legend).}
\end{figure}

Figure \ref{fig:CloudConv1}(a) shows the $3D$ chart in the parameter space containing all $300$ individuals involved in the optimization process. Data points distributed all over the volume are typically generated by the random stage 1 ($m<p_{th}=50$) and GOs tend to accumulate solutions around small volumes where fitness is typically small, with the overall effect of monotonically decreasing the average fitness value, observed when fitness is represented in order of execution [black curve in Fig. \ref{fig:CloudConv1}(b)]. However, the scattering ability of GOs often results in finding slightly better individuals in nearby regions presenting smaller agglomeration. An important reason for the convergence of our GA towards the optimal solutions is the fact that the operator $\hat{\mathpzc{M}}$ is given a lower probability of action than $\hat{\mathpzc{X}}$ (probabilities are $1/3$ and $0.95$ respectively, see previous section). This combination gives both a good diversity and  probability to conserve the properties of the best individuals during the execution of the GA.
%FIG 4

\section{Conclusions\label{sec:conc}}
We presented an efficient optimization procedure based on GAs deployed in the Grid platform, providing faster results and potential scalability of the computational resources. The optimization provides the optimum input pulse parameters required to control the SC dynamics in a way that the first two ejected Raman solitons are centered at two pre-defined wavelengths. The results are shown to be of interest for practical OCT applications in the NIR II region where dual frequency, pulsed sources enable in vivo imaging, and avoid spurious results.

%\section*{Acknowledgements}
%F.R.A.S. thanks the Consejo Nacional de Ciencia y Tecnolog\'ia (CONACyT). F.R.A.S. and  M.T.C. acknowledge partial funding provided by the projects CONCyTEG (GTO-2012-C03-195247) and DAIP-UG  382/2014. I.T.G. acknowledges CONACyT for partial support, project: 106764 (CB-2008-1). The work of A.F. was supported by the MINECO under Grant No. TEC2010-15327. C.M. thanks Dr. Miguel Arevalillo Herr\'aez for details on GAs. \\

%This paper was published in Opt. Express and is made available as an electronic reprint with the permission of OSA. The paper can be found at the following URL on the OSA website: \url{http://www.opticsinfobase.org/oe}. Systematic or multiple reproduction or distribution to multiple locations via electronic or other means is prohibited and is subject to penalties under law.
\end{document}